\documentclass[12pt,a4paper,twoside]{article}

\usepackage{graphicx}
\usepackage{epic}
\usepackage{epstopdf}
\usepackage{amssymb}
\usepackage{amsmath}
\usepackage{sectsty}

\usepackage[utf8x]{inputenc} 
\usepackage[english]{babel}

\sectionfont{\huge}

\begin{document}

\begin{center}
{\LARGE\bf Optimized Integral Controller Searching Prime Number Orders}
\end{center}

\vspace{0.1cm}
\begin{center}
{\it Petr Kl\'{a}n, Dept. of System Analysis, University of Economics in Prague, Czech Republic, petr.klan@vse.cz}
\end{center}

\vspace{0.7cm}
\noindent
{\bf Abstract:} The concept of integration is generally applicable to automatic control of processes. As shown in this paper, integral controller performs efficient searches in the extensive prime sets, too. An inspiration by the simple analytic rules for PID controller tuning results in integral controller that ensures predictable work regardless of the cardinality of primes. It gives an innovative application of the feedback control which relates to a gradient method. 

\vspace{1cm}
\noindent
{\bf Keywords:} PID control; Integral controller; Controller tuning; Static process; Table of primes

\section{Introduction}

\vspace{0.3cm}\noindent Integral (shortened by I) part of PID (Proportional plus Integral plus Derivative) control of dynamic processes is one of the most frequently used components due to its ability to eliminate control errors \cite{sem}. Much attention is therefore focused to its analytic tuning  as shown e.g. in \cite{odw}. Simultaneously, I algorithm represents a dynamic process model. If I acts in the stage of a PID controller, it is possible to state that one dynamic process maintains another dynamic process. But can a dynamic process adjust a static process, which is the opposite of dynamic process?\\
\\
Conversely, it is possible partially. It is the case where a static process (P controller) maintains the dynamic process. For non--integrating processes it is known that P controller decreases control errors, a complete elimination of the latters is possible only in singular cases of infinite proportional gains \cite{kg1}. A similar function is performed by the so--called signal followers. Conversely, when I controller is used in adjusting of a static process, errors can be quickly eliminated due to integration effects as it can be easily derived.\\   
\\
A prime number is a natural number greater than $1$, that has no divisors others than $1$ and itself \cite{cogu}. For example, number $17$ is the prime because it has only two divisors: numbers $1$ and $17$. Tables of primes are very frequent in history of numbers. Online tables on {\it The Prime Pages} incorporate the first $50$ millions of primes as well as many research results in the subject of prime number theory. In this paper, a process with the first $30$ millions of primes will be used. Considering the prime process, the following objective will be formulated: design a controller such that for an arbitrarily selected prime a corresponding index--order will be determined via a good control manner. The latter includes fast searches with predictable number of steps at minimum changes consumed by searches \cite{mmi}.\\
\\
The static prime process is introduced and some properties from control point of view are established in the following. This process is represented by a single block having one input (natural index--order) and one output (associated prime). A concept of the feedback loop is then formed with presence of I controller in an incremental form. Adaptive tuning of this controller is proposed. It results in the effective searching of primes by the control of order indexes associated with these primes.   

\section{Prime Process}   

\begin{figure}[htb]
\centering
\includegraphics[scale=0.7]{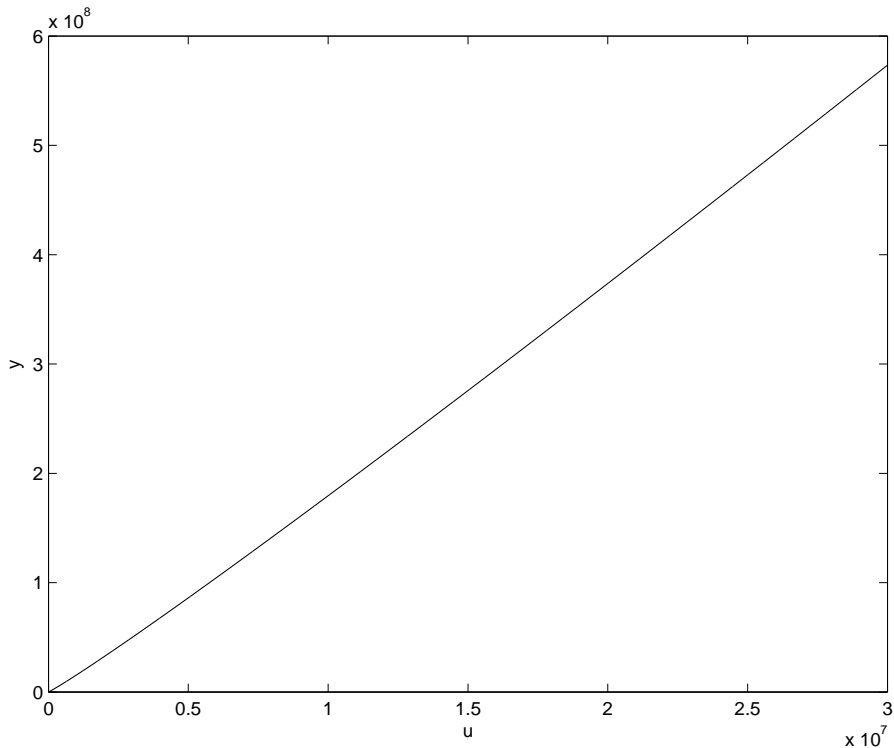}
\caption{\small Static characteristics of the prime process.}
\label{statchar}
\end{figure}

\begin{figure}[htb]
\centering
\includegraphics[scale=0.7]{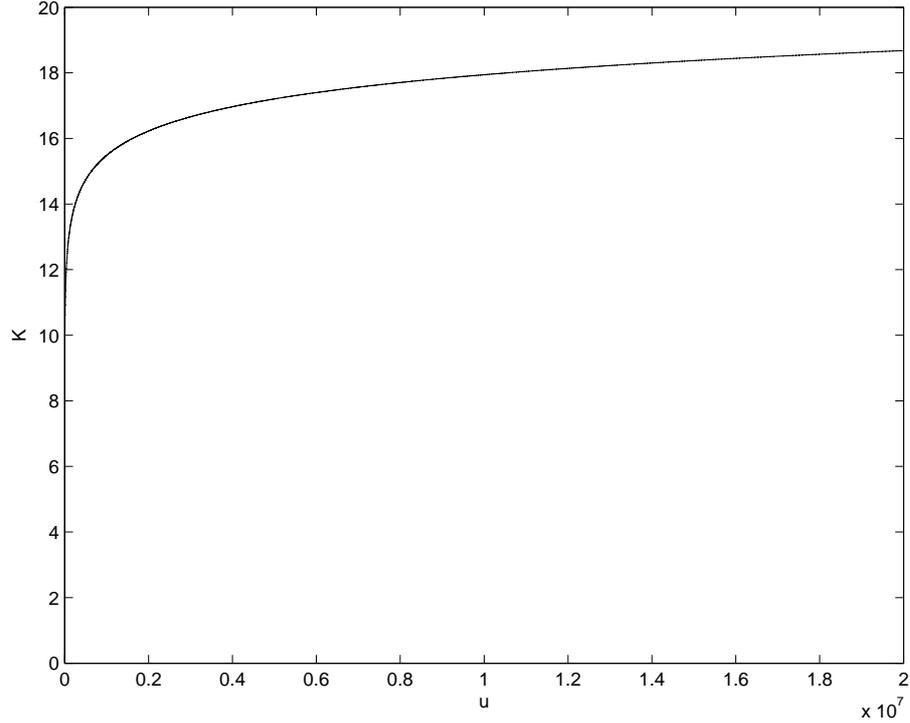}
\caption{\small Gain of the prime process.}
\label{gain}
\end{figure}

\vspace{0.3cm}\noindent In order to form a static prime process, prime numbers are arranged into a countable array 

\begin{equation}
y=p(u), \label{primeprocess}
\end{equation}
where $u$ is an input index indicating prime order from beginning ($u = 1,2, \ldots, 30\,000\,000$) and $y$ is $u$--th prime number. Obviously, $2=p(1)$, $29=p(10)$, $541=p(100)$, $7\,919=p(1\,000)$, $15\,485\,863=p(1\,000\,000)$ etc. Relation (\ref{primeprocess}) is called {\it prime process}. This process assigns some prime number to each natural index--order $u$, i.e. a one--to--one correspondence between the finite set of natural order indexes and the finite set of primes is set up.\\
\\
In the automatic control, it is usual to first deal with the process gain $K$. A quick view of the static characteristics of the prime process given in Fig. \ref{statchar} indicates that it corresponds to a nearly linear process. An analysis of the values above, however, maintains a distinctive style, since $ p(1) / 1 = 2$, $p(10) / 10 = 2.9$, $p(100) / 100 = 5.41 $, $p(1\,000) / 1 \, 000 = 7.919$, $p(1 \, 000 \, 000) / 1 \, 000 \, 000 = 15.485863$ etc. Prime process gain grows steadily even though from one million of primes very slightly. Actual gain for the first $20$ millions of primes is shown in Fig. \ref{gain}. The latter confirms a nonlinear dependence specifically in the process beginning. Note, that the average gain of the prime process approximately is $K =  18$.

\vspace{0.3cm}\noindent From the theory of prime numbers it is known \cite{cogu} that for $u$--th prime number $p (u)$ is $p(u) \sim u\ln u$. Hence, the relation $K(u) \sim \ln u$ is easily obtained for the corresponding actual gain. Notation $\sim$ denotes asymptotic equality, i.e. $\lim_ {u \rightarrow \infty} K (u) / \ln u = 1$. In other words, for a sufficiently large $u$, there is possible to substitute $K (u)$ by $\ln u $. In the case of $u = 30 \, 000 \, 000$, $ K (u) = p (30 \, 000 \, 000) / 30 \, 000 \, 000 = 19.1$ and $ \ln (30 \, 000 \, 000) = 17.2$ which indicates $10\%$ error. According to the updated asymptotic estimate $ p (u) \sim u  \ln u + u [\ln \ln u-1] $ is $ K (u) \sim \ln u + \ln \ln u  -1$ which relates to the refined estimate of $ K (30 \, 000 \, 000) $ in the form $ \ln (30 \, 000 \, 000) + \ln \ln (30 \, 000 \, 000) -1 = 19.06$. The latter nearly presents the gain accurate value.  \\  
\\
Note that time constants of processes represent another common parameter. Prime process is static, in that case, therefore, the time constant stays zero.

\section{Use of I algorithm}   

\begin{figure}[hbt]\begin{center}
\setlength{\unitlength}{1mm}

\begin{picture}(155,80)
%Process and controller
\thicklines \put(35,40){\framebox(23,15){I controller}}
\put(92,40){\framebox(23,15){$p(u)$}} \thinlines

\thicklines \put(21,47.5){\circle{8}}  \thinlines

\put(1,47.5){\vector(1,0){16}}
\put(25,47.5){\vector(1,0){10}}\put(58,47.5){\vector(1,0){34}}
\put(115,47.5){\vector(1,0){40}}

\put(1,49.5){\makebox(0,0)[l]{\small $w$}}
\put(67,49.5){\makebox(0,0)[l]{\small $u$}}
\put(149,49.5){\makebox(0,0){\small $y$}}
\put(13.5,49.5){\makebox(0,0)[l]{\small $+$}}
\put(16,41.5){\makebox(0,0)[l]{\small $-$}}

\put(141,47.5){\line(0,-1){27.5}}\put(141,47.5){\circle*{0.5}}
\put(21,20){\vector(0,1){23.5}} \put(21,20){\line(1,0){120}}
\end{picture}
\caption{Prime searches via feedback organization.} \label{fb}
\end{center}
\end{figure}
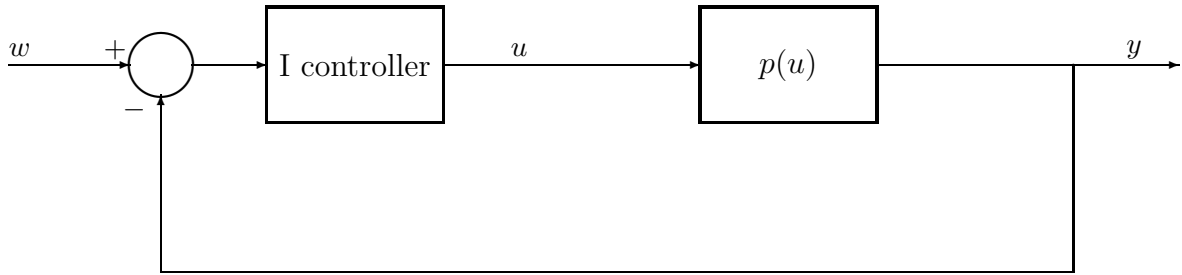

\vspace{0.3cm}\noindent Consider a preselected prime number $y$. The objective is to find a prime process input $u$ associated with the prime $y$. In order to show that it is not a trivial task, try out a series of manual searches. Consider more carefully the preselected prime e.g. $ y = 141 \, 661 \ 147 = p (u=?)$ where the input index is not a priori known. Manual control of the trained author can be represented by the following sequence of $10$ steps (initial state is determined by $ u (0) = 1 $ and $ y (0) = 2 $): 

\begin{itemize}
\item $u(1)=10\,000\,000$, $y(1)=179\,424\,673$ 
\item $u(2)=9\,000\,000$, $y(2)=160\,481\,183$
\item $u(3)=8\,000\,000$, $y(3)=141\,650\,939$
\item $u(4)=8\,000\,200$, $y(4)=141\,654\,581$
\item $u(5)=8\,000\,400$, $y(5)=141\,658\,373$
\item $u(6)=8\,000\,500$, $y(6)=141\,660\,191$
\item $u(7)=8\,000\,550$, $y(7)=141\,661\,081$
\item $u(8)=8\,000\,553$, $y(8)=141\,661\,129$
\item $u(9)=8\,000\,554$, $y(9)=141\,661\,139$
\item $u(10)=8\,000\,555$, $y(10)=141\,661\,147$
\end{itemize}  
Search sequence shows that the selected prime is $ 8 \, 000 \,  555$--th in the order from beginning. By other formulation $ 141 \, 661 \ 147 = p (8 \, 000 \, 555) $ or $8 \, 000 \, 555  = p^{-1} (141 \, 661 \ 147) $, where $p^{-1}(.)$ denotes inverse of prime process (\ref{primeprocess}). Manual search is arduous since comparing large numbers and it very much depends on the estimate for the first order $u(1) $. Clearly, there can be used information about average gain in the prime process $ K = 18$. However, it does not reduce arduous work when comparing large numbers.\\
\\
Basic feedback layout for the automatic solution of similar task is shown in Fig. \ref{fb}.  Schematically, the latter is equivalent to the classical feedback control incorporating a process and a controller. Set--point $w$ represents the preselected prime number whose index--order is requested to determine. Output of integral controller immediately acts on prime index--orders which results in a final balance determined by $y = w$.

\vspace{0.3cm}\noindent Since the output of the I controller can generally produce real numbers and the prime process input considers natural numbers only, the controller output is adapted by rounding operation known as $u = floor(x)$ where $x$ is rounded to the nearest natural number lower or equal to $ x $. For example, $ 3 = floor (3.71) $ etc. Currently, $floor$ operations are routinely denoted by $ \lfloor x \rfloor$. \\
\\
In this paper, integral control is based on the continuous--like case \cite{sem}. If the I control is specifically represented by transfer function $ K_c / (T_i s) $ where $K_c$ is a controller gain and $T_i$ an integration time constant then the transfer function of the feedback arrangement according to Fig. \ref {fb} will be

$$
\dfrac{\dfrac{K}{T_i s}}{1+\dfrac{K}{T_i s}}=\dfrac{1}{\dfrac{T_i}{K}s+1}
$$ 
with the resulting closed--loop gain $1$ and the possibility of accelerating searches of prime numbers by decreasing integration constant $T_i $  (under simplification of $ K_c = 1 $). \\ 
\\
Integral algorithm is used in a conventional incremental form

$$
u(k+1)=u(k)+\frac{1}{T_i} e(k),
$$
where $ e (k) = w-y (k) $ denotes control error and $ k $ denotes step. If the input of the prime process is $ \lfloor u (k) \rfloor $, then the related output is $y (k) = p (\lfloor u (k) \rfloor) $ for each step $ k $. Searches finish when $ y (k) = w $ are equable. Here, the final controller output $\lfloor u (k)\rfloor$ determines the natural order of the preselected prime set--point. In Matlab (similarly Scilab), it associates source code in each step:

\begin{verbatim}
y=p(floor(u));
e=w-y;
u=u+(1/Ti)*e;
\end{verbatim}
When an overflow outlet of the integral controller is indicated, it is requested to disable integration by activating an antiwindup technique. Single antiwindup restrictions are provided for the set of $30$ millions primes as follows:

\begin{verbatim}
if (u>30000000) u=30000000; end
if (u<1) u=1; end
\end{verbatim}
   
\section{Tuning of I algorithm}   

\begin{figure}[htbp]
\begin{center}
\includegraphics[scale=0.7]{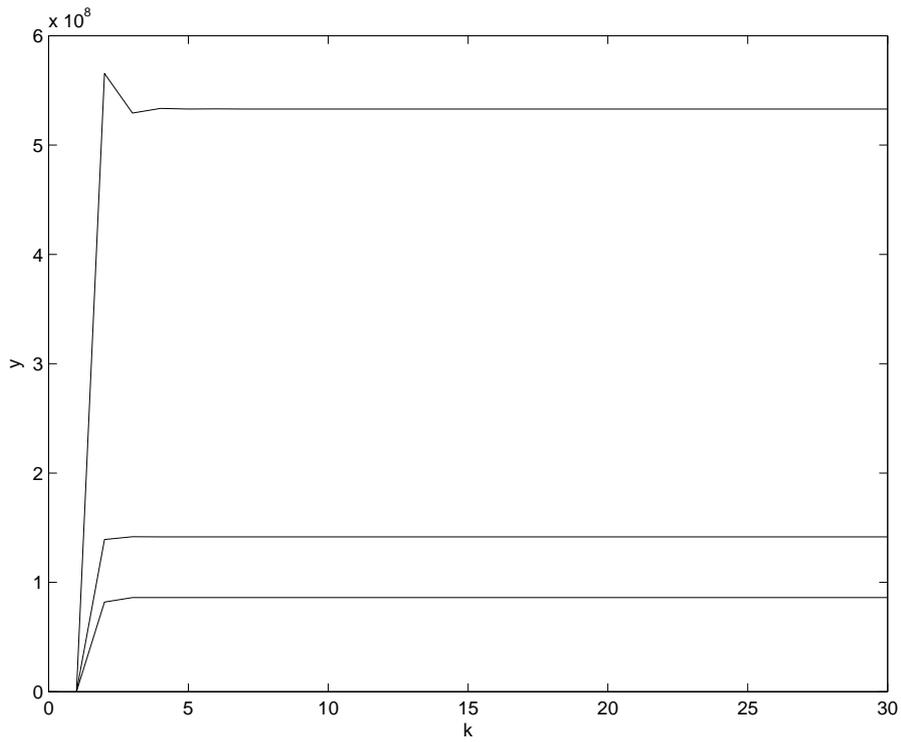}
\end{center}
\caption{\small Automatic prime searches $y$ for $w=86\,028\,121$, $w=141\,650\,939$ and $w=533\,000\,389$.}
\label{primesearch}
\end{figure}

\begin{figure}[htbp]
\begin{center}
\includegraphics[scale=0.7]{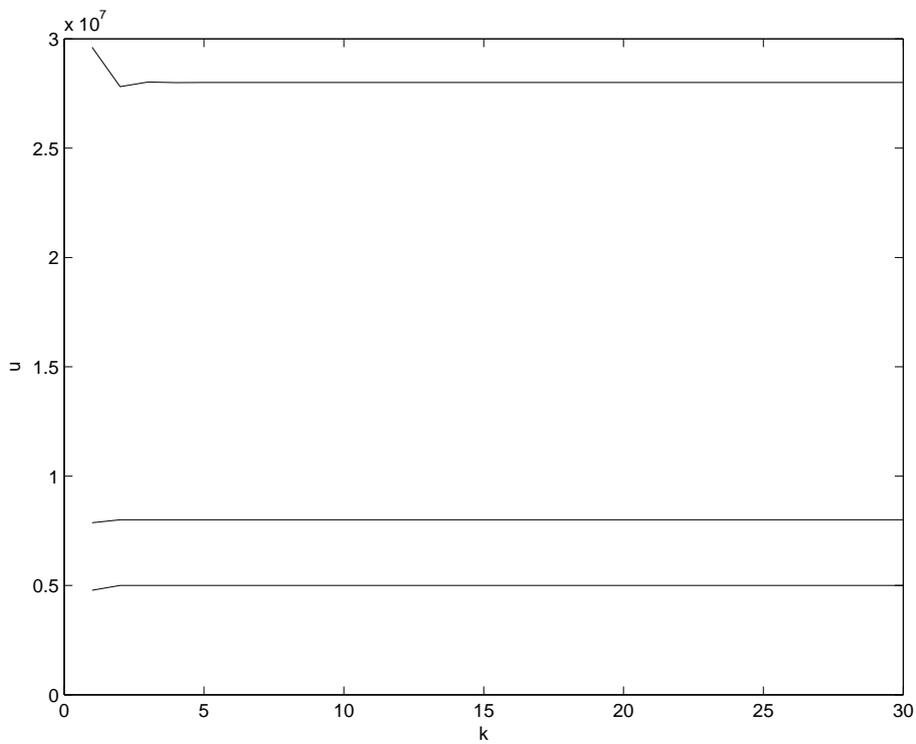}
\end{center}
\caption{\small Prime orders $u$ in prime searches for $w=86\,028\,121$, $w=141\,650\,939$ and $w=533\,000\,389$.}
\label{primesearchu}
\end{figure}

\vspace{0.3cm}\noindent A natural setting of the integral time constant (associated with so called balanced tuning of PI controller \cite{kg2} is 

$$T_i=K.$$ 
This tuning ensures finding of prime orders practically in three steps, as it is verified in Fig. \ref{primesearch} for various preselected primes: $ w = 86 \, 028 \,  121$, $ w = 141 \ 650 \,  939$ and $ w = 533 \, 000 \,  389$. Horizontal axis represents the steps of $k$ movement. Searching dynamics of course varies due to changes of the actual prime process gain. Corresponding sequence of natural prime orders $ u $ is in Fig. \ref {primesearchu}. In comparison with manual searches of the author above, more than three times acceleration in search speed is observed. Moreover, these searches are predictable with regard to the number of steps, which is almost unattainable in the case of manual searches.

\vspace{0.3cm}\noindent Due to the aforementioned prime theorem estimation $p(u) \sim u\ln u$ and in the case of sufficiently large primes, tuning of the integral time constant is preferred in the form 

$$T_i=\ln w,$$ 
where $ w = p (u=?)$ is the $u$--th known prime set--point. Since $ \ln w = \ln u + \ln \ln u $ which presents an estimate of $K (u) $ plus some small number given by a very few growing double logarithm, the setting of the integral time constant is a few higher than the actual gain is. It improves safety of I control against instability while slowing down the prime searches very slightly.

\section{Simple Proportion--Based Controller}

\vspace{0.3cm}\noindent The integration algorithm $u(k + 1) = u (k) + e (k) /T_i$ can be replaced by the following simple proportion--based way. Considering the prime process approximate relation $y (k) = Ku(k) $, one can make the following deduction 

$$
\frac{u(k+1)-u(k)}{u(k)}=\frac{K}{T_i}\frac{e(k)}{y(k)}=\frac{w-y(k)}{y(k)}.
$$
Hence ($K=T_i$)

$$
u(k+1)=u(k)\frac{w}{y(k)}=u(k)\frac{w}{p(\lfloor u(k)\rfloor)}
$$
represents an iterative formula for finding the prime order based on the use of simple proportion $w/u(k+1)=y(k)/u(k)$. It is associated with an one--step ahead Matlab iteration 

 \begin{verbatim}
 y=p(floor(u));
 u=u*w/y;
 \end{verbatim} 
There is the one iteration only in the case of a purely linear relationship $y = Ku$, since $y (k + 1) = Ku(k+1) = Ku(k) w / y (k) = w $. Experiments show that in the case of nonlinear prime process iterations converge very quickly in a few steps.

\section{Conclusion}

\vspace{0.3cm}\noindent The paper shows how it is possible to use the integration algorithm to solve the problem so far away with respect to the current control, such as automatic searches in prime tables. Feedback arrangement with use of a pure integral controller guarantees that searches work efficiently with a minimal amount of energy. It is associated with a good controller tuning. \\
\\
Illustratively, there is shown the benefit of automatic searches, when a very simple algorithm, unlike the manual search, provides predictable results regardless of the size of the prime numbers and the extent used sizes.\\
\\
As illustrated above, the very simple feedback procedure yields surprisingly good results. In fact, it gives an alternative to feedback scheme proposed in \cite{ggs} aimed at finding the inverse of a general function or an alternative to searches of measurable records in relational databases.\\
\\  
In prime numbers in general, apparently experimental evidence often carries much less weight than it seems. An open question is whether proposed I algorithm applies to any prime outside the first $30$ million table.

\end{document}